# Small Bodies of the Solar System Active at Large Heliocentric Distances: Studies with the 6-Meter Telescope of SAO RAS

A. V. Ivanova[1, 2, 3]

[1]Astronomical Institute of the Slovak Academy of Science, Tatranska Lomnica, SK-05960 Slovak Republic
[2]Main Astronomical Observatory of the National Academy of Science of Ukraine, Kyiv, 03143 Ukraine
[3]Astronomical Observatory of the Taras Shevchenko National University of Kyiv, Kyiv, 04053 Ukraine

A detailed study of comets active at large heliocentric distances (greater than 4 au) which enter the Solar System for the first time and are composed of matter in its elementary, unprocessed state, would help in our understanding of the history and evolution of the Solar System. In particular, contemporary giant planet formation models require the presence of accretion of volatile elements such as neon, argon, krypton, xenon and others, which initially could not survive at the distances where giant planets were formed. Nevertheless, the volatile components could be effectively delivered by the Kuiper-belt and Oort-cloud bodies, which were formed at temperatures below 30 K. This review is dedicated to the results of a multi- year comprehensive study of small bodies of the Solar System showing a comet-like activity at large heliocentric distances. The data were obtained from observations with the 6-meter telescope of SAO RAS equipped with multi-mode focal reducers SCORPIO and SCORPIO-2.

## I. INTRODUCTION

Contemporary studies of small Solar System bodies (hereafter SSSB) involve various aspects related to their formation in the protoplanetary disc, their dynamical and chemical evolution, ac- tivity depending on the distance from the Sun, as well as investigating their physical and dynamical characteristics, including their interconnection. Recently, new data have been obtained related to the physics of the small Solar System bodies, which changed the conceptions about the formation and evolution of the Solar System and the interrelation between different populations of small bodies. Thus, based on the latest studies Dones et al. [19], one should assume the existence of three reservoirs of comets in the Solar System: the trans-Neptunian region, the Oort cloud, and the main asteroid belt (Fig. 1).

Dynamical studies of the SSSB indicate the Oort cloud as the main source of *nearly isotropic comets*, including long period comets (LPC) and Halley-type comets. In turn, ecliptic comets (EC) and Encke-type Comets (ETC) were formed mainly in the Kuiper belt, and the scattered disc is the source of the Jupiter-family comets (JFC) and centaurs [69, 118, 119]. However, these and other studies of the Solar System body dynamics [19, 20, 68, 90] did not provide an answer to the question of whether the short- and LPCs formed in different regions of the Solar System or in the slightly overlapping regions of the initial protoplanetary disc before being ejected into the outer regions of the Solar System.

Active SSSB studies in recent years lead to an accumulation of large amounts of data on short- period (ShPC) and LPC at small (less than 2 au) distances from the Sun. On the other hand, comets with a perihelion greater than 4 au were rarely observed, leading to a significant lack of information on the comets that are active at large distances from the Sun. This situation resulted in the question of the comet taxonomy also remaining open. Up to now, groups of comets could not be classified based on their physical and chemical characteristics, with the exception of a group of comets with a carbon compound deficiency. However, most of them are ShPC, and this feature may be related to the region of their formation in the Solar System and not their evolution [11, 77].

---

*Electronic address: oivanova@ta3.sk

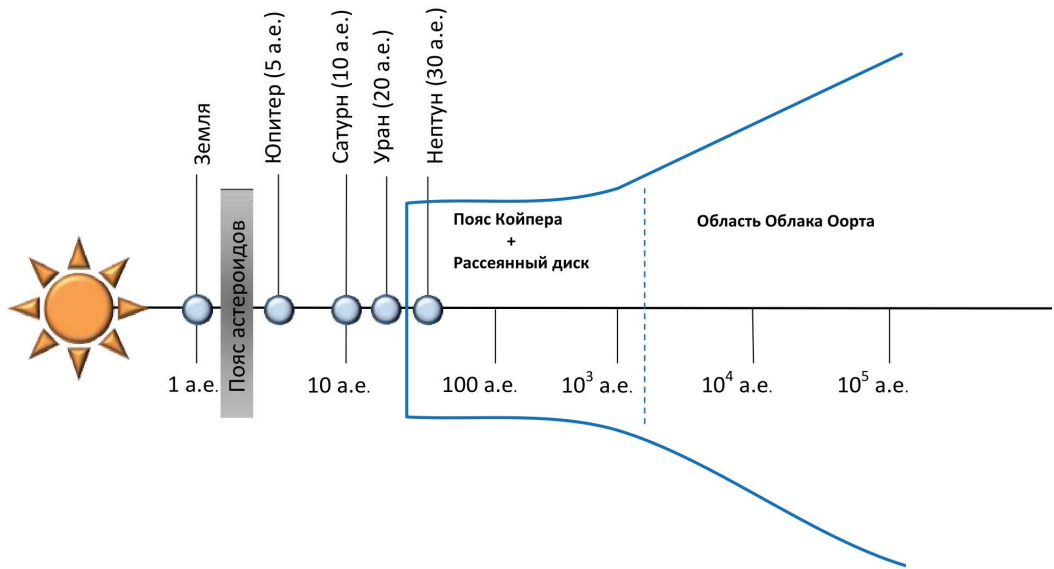

Figure 1: Schematic image showing the Kuiper belt, the scattered disc and the Oort cloud to scale with our planetary system.

Owing to the space missions to several comets[1], new and unique data were obtained on the physical and chemical properties of comet nuclei and cometary dust. It was established that:

- the surfaces of the studied comets vary significantly both in structure and composition;

- most of the surfaces are covered by a dark dusty crust, and the ices are present in the form of local spots;

- matter ejection in jets does not stop on the night side of the comet nuclei;

- comet activity may be regulated not only by the water ice sublimation, but also by sublimation of more volatile nuclei components such as CO and $CO_2$;

- comet dust may contain both the protosolar nebula and interstellar particles.

However, space missions cannot be very frequent, and the investigated objects are limited to EC (with the exception of a single comet, 1P/Halley, which belongs to the group of nearly isotropic comets). Only in 2019 did ESA support for the first time the new "Comet Interceptor" mission to a LPC within the framework of the "Cosmic Vision" program[2]. Therefore, ground-based observations and monitoring of comets active at large heliocentric distances remain very relevant and important for comet physics.

---

[1] https://nssdc.gsfc.nasa.gov/planetary/planets/cometpage.html

In 2006 our group started a comprehensive program to study the activity of comets at large heliocentric distances using the 6-meter telescope of the Special Astrophysical Observatory of the Russian Academy of Sciences (SAO RAS). Due to the uniquely designed SCORPIO [2] and SCORPIO-2 [1, 3] instruments, we were able to conduct detailed studies of the SSSBs using var- ious techniques. In particular, we were able to obtain spectral, photometric, polarimetric and interferometric data for a selected object on the same night with one telescope.

The main program tasks were:

- determining the physical properties of selected SSSB objects belonging to different dynamic groups;

- investigating the manifestations and development of non-stationary processes in objects with detected activity;

- searching for molecular emissions in the spectra of distant comets, responsible for their activity and ejection of matter from the surface;

- investigating the evolution of activity as a function of heliocentric distance for new comets;

- studying the dust component in the atmospheres of comets that enter the inner regions of the Solar System for the first time.

Such an approach, as well as using a large telescope, allowed us to broaden our knowledge in the comet physics field. The data obtained over the course of ten years opened new possibilities for the interpretation of the activity of distant active comets [42, 43, 57–61, 64, 102], ionization at large distances from the Sun [39], composition and shape of the particles forming the comet atmosphere [17, 42, 48], and allowed us to introduce model parameter restrictions [17, 42]. Further analysis of the accumulated data will, in time, allow us to determine possible dynamical connections between various SSSB populations by means of their orbital evolution modeling, and the relation between the physical and dynamical characteristics of SSSBs from different populations in order to reveal the features related to different regions of their formation and evolution.

In this paper we present a series of unique results of our research program, which allowed us to find explanations for several unresolved comet physics problems, and introduce new tasks for further investigation.

## II.   SPECTRAL STUDIES OF DISTANT COMETS

Spectral studies reveal information about the neutral ionized gas, as well as dust, which are present in comet atmospheres. An analysis of the emission from both these components is the main source of information about the physical processes in the atmospheres of comets. The physical parameters obtained for the gas and dust components allow us to develop a quantitative theory of comet atmosphere emission. Spectrophotometric comet studies are very informative, however, unlike photometry, high quality comet spectra (especially for distant comets) can only be obtained using large telescopes. In 2006, the 6-m SAO RAS telescope started a comprehensive program aimed at investigating the activity of distant comets, including a search for gas emissions responsible for the formation of comets and the tails of these comets. Long-slit spectra were obtained for more than 30 distant comets over the course of ten years. This allowed us to organize and broaden our knowledge on LPCs, since only a few of them had such data available previously. An analysis has shown that most of the considered comets demonstrate only continuous spectra [42, 43, 60, 102], determined by the scatter of solar radiation on comet dust (Fig. 2).

_______
[2] http://www.esa.int/Our_Activities/Space_Science/ESA_s_new_mission_to_intercept_a_comet

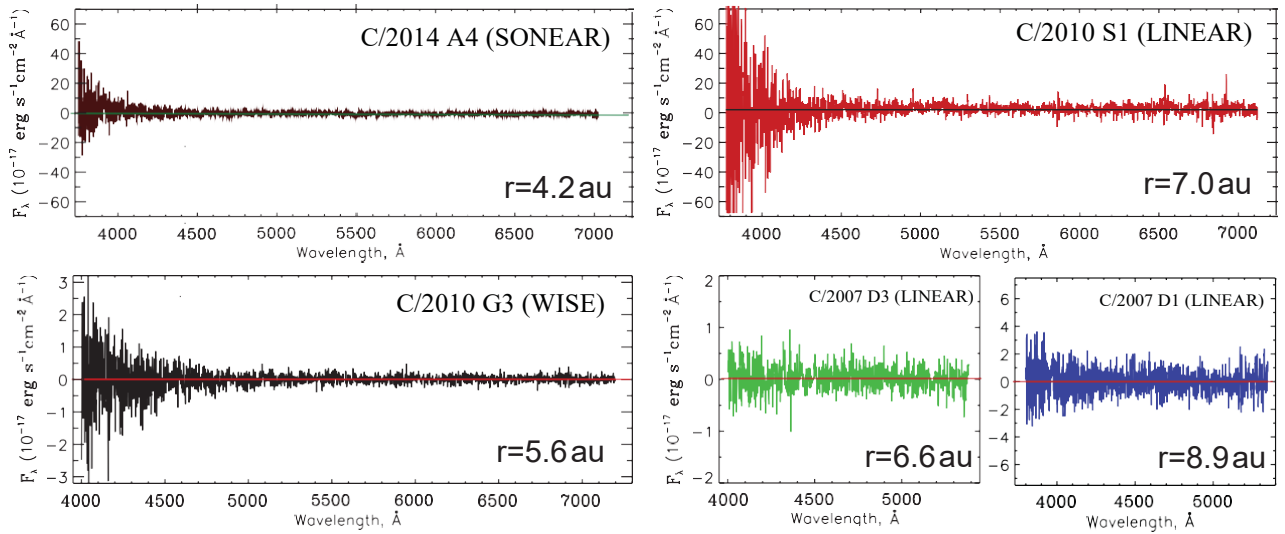

Figure 2: Spectra of five distant comets [42, 43, 45].

Wavelength dependences of dust scatter efficiency were obtained for distant comet spectra. In most cases this allowed us to register the reddening effect (scatter efficiency increasing with wavelength) [42, 43, 49, 57, 58, 60, 61]. However, for some comets the efficiency of solar emission scatter by dust was higher in the blue spectral region than in the red [60].

Significant emission components were detected only for three comets: 29P/Schwassmann– Wachmann 1 (hereafter 29P), C/2002 VQ94 (LINEAR) (hereafter 2002 VQ94), and C/2011 J1 (Catalina) (hereafter 2011 J1) [46, 57, 58]. Here we identify CN (29P, 2002 VQ94, 2011 J1) [46, 57, 58] and $C_3$ (2002 VQ94) [57, 58] neutral molecule emissions, and also rather strong $CO^+$ and $N^+$ ion emission bands in the spectra of comets 2002 VQ94 and 29P [57, 58, 61]. The presence of $CO^+$ and $N^+$ in the spectra of 29P was later confirmed by the SOAR telescope observations [49, 50]. Prior to our observations, the $CO^+$ ion emissions were registered only in the spectra of comet 29P, at distances of the order of 6 au from the Sun in the 1980's. The detection of $N_2$ in comets was a matter of dispute for many years, since the $N_2$ molecule, due to its physical properties, cannot be detected in the visual range of the comet spectra in ground-based optical observations. Before the publication of our studies of distant comets [49, 50, 57, 58, 61] the $N^+$ ion was observed in the spectra of several comets located close to the Sun: C/1908 R1 (Morehouse) [15], C/1961 R1 (Hu- mason) [24], 1P/Halley [75, 122], C/1987 P1 (Bradfield) [75] and C/2016 R2 (PanSTARRS) [12]. The $N_2$ molecule was first discovered in the 67P/Churyumov–Gerasimenko comet (hereafter 67P) by the ROSINA mass spectrometer within the framework of the Rosetta space mission [103].

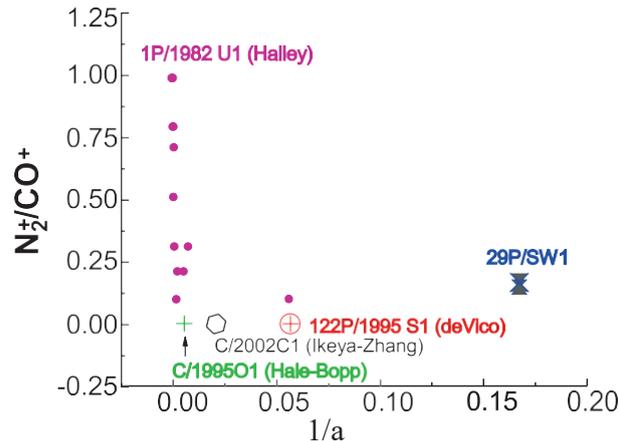

Figure 3: Estimate of the $N^+(0, 0)/CO^+(4, 0)$ ratio for different dynamical types of comets. $a$—semi-major axis. The comet data were taken from the following papers: 29P/Scwassmann–Wachmann 1—Ivanova et al. [49]; 1P/Halley—Cochran et al. [10]; 122P/de Vico, C/2002 C1 (Ikeya–Zhang), C/1995 O1 (Hale–Bopp)— Cochran [8], Cochran et al. [10].

Spectral monitoring of comet 2002 VQ94 has shown that ions $CO^+$ and $N^+$ in the comet spectra are also observed at heliocentric distances of 7.33 and 8.36 au [61]. Only at a heliocentric distance of 9.86 au have we stopped detecting the emissions of these ions, which, possibly, has to do with both the absence of emissions and with the instrument sensitivity and, correspondingly, with the weak signal from the comet at such large heliocentric distances.

It was shown back in the 80's that CO, $H_2O$ and $N_2$ should be the dominating equilibrium types of carbon, oxygen and nitrogen in the early solar nebula conditions [70]. Therefore, studying the nitrogen abundance in comets is important for understanding of the conditions under which they were formed in the early solar nebula. It is generally accepted that CO- and $CO_2$-ices were the most common types of ice in the protoplanetary nebula (along with water ice) [86]. The ions $CO^+$ and $N^+$ which we identified in distant comets, where pristine protoplanetary nebula material remains intact, confirm this conclusion. The upper limits of the $N_2/CO$ ratio for several comets and the reason for the $N_2$ deficiency in comets is discussed by Cochran et al. [10]. Our estimates for the $N^+/CO^+$ ratio for selected distant comets, as well as results presented in the literature, are shown in Fig. 3.

In addition to individual spectral observations of selected distant comets, we investigated the parameter variations of comet spectra as a function of distance to the Sun. For several comets (for example, C/2006 W3 (Christensen) and C/2012 S1 (ISON)), which were observed on large heliocentric distances, we discovered that well-developed emission spectra are formed at distances less than 4 au from the Sun. For instance, neutral molecule and ion emissions have not been registered in the spectra of comet C/2006 W3 (Christensen) when it was located in the 8.49 to

6.25 au heliocentric distance range. The comet spectrum changed significantly after it reached

3.7 au from the Sun. Emissions of CN, $C_3$, $C_2$, CH, $NH_2$ molecules were then identified there, as well as those of $CO^+$ and $CH^+$ ions [59]. A similarly rich emission spectrum was also observed for the unique comet C/1995 O1 (Hale–Bopp) at small distances from the Sun. As is the case for distant comets 2002 VQ94 and 29P, multiple $CO^+$ emissions have been identified for the comet 2006 W3, and this could indicate that the comet is enriched with CO.

Based on the results of spectral studies, it can be said that a fraction of the distant comets that we considered are enriched with CO and $N_2$, and that can be directly related to the place of formation of these comets in the Solar System. Additionally, our results confirm the idea that volatile matter is delivered when giant planets are formed.

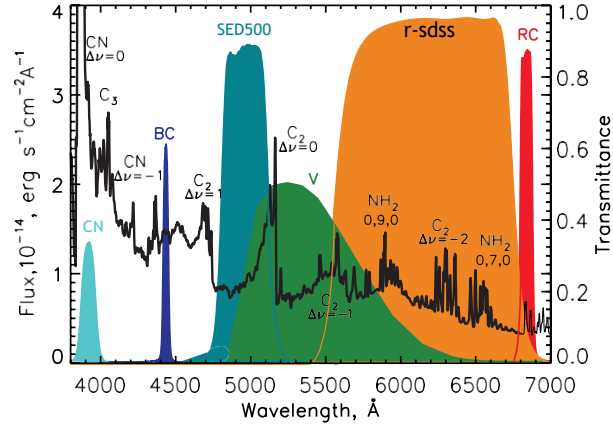

Figure 4: Normalized transmission curves of filters *CN* ($\lambda$ 3982/36 Å), *BC* ($\lambda$ 4429/36 Å), *RC* ($\lambda$ 6835/83 Å), *SED*500 ($\lambda$ 5019/246 Å), *r — sdss* ($\lambda$ 6200/4200 Å) and *V* ($\lambda$5580/880 Å), superimposed on the spectrum of comet 2P/Encke, obtained with the 6-m telescope of SAO RAS.

## III.    PHOTOMETRY OF DISTANT COMETS

Among the methods used for determining the physical properties of comets, photometry is the most advanced and widespread. Photometric observations of comets are often conducted with broadband filters, which in our case are the Johnson–Cousins *UBVR_C* and SDSS *ugriz* systems. Unfortunately, the main comet emissions CN, $C_2$ and $C_3$ do not coincide with the transmission peaks of the standard broadband filters (Fig. 4). Therefore, the color indices of those comets whose spectra exhibit strong gas emissions, obtained using those filters, do not adequately represent the properties of comet dust. Nevertheless, such observations are useful both for studying the dust properties of comets active at distances higher than 4 au [42, 43, 48, 59, 60], where strong spectral emissions are for the most part absent, (see Fig. 2), and for investigating active processes in comet atmospheres, estimating the volume luminosity and density of matter in the head of the comet, as well as determining the rotation period of the nucleus, among other tasks.

For distant comets, the spectra of which exhibit gas emissions [57, 58], broadband photometry was supplemented with spectral data, which allowed us to improve significantly the interpretation of broadband photometric observations and improve the accuracy of the determined physical prop- erties. Narrow band filters developed within the framework of the C/1995 O1 (Hale–Bopp) comet international research program were used for photometric tasks related to studies of the gaseous component (*CN* band observations of individual molecular emissions of some distant comets), CO$^+$ ion coma, and the dust component in comet continuum regions [22]. These filters are meant to isolate individual molecular emissions and regions in the comet continuum. In the 6-m SAO RAS telescope observations, a set of filters with a 250 Å band width (SED) was also used for such tasks, manufactured by Edmund Optics[3].

Despite the short comet visibility time periods, our episodic but well-planned observations allowed us to obtain rich material on the physical characteristics of these unique objects in the Solar System, combining the data obtained by various methods. Thus, carrying out our comprehensive program aimed at the study of the activity of SSSBs at large heliocentric distances using the 6-m telescope of SAO RAS enabled us to determine from photometric data the gas and dust ejection velocity (as well as their variations with heliocentric distance), a comparative chemical composition of the comets, and estimate the rotation periods of their nuclei, their sizes, and much else.

---
[3] https://www.sao.ru/Doc-k8/Telescopes/bta/instrum/

However, long series of systematic observations are required to solve some problems of comet physics, which are impossible to obtain using large telescopes due to the limited allocated observing time. In this case, we conducted parallel monitoring with various small and medium-sized telescopes [40, 41].

Over the ten years of photometric SSSB observations with high spatial resolution using the 6-m SAO RAS telescope, more than 45 comets of different dynamic groups were investigated, as well as individual active asteroids and centaurs.

Special attention was paid to those SSSBs which are active at significant heliocentric distances, the majority of which have perihelion distances greater than 4 au. The question of comet activity evolution as it approaches the Sun from distances exceeding 10 au was studied separately [102].

Particular attention was also devoted to the task of comparing the changes in the dust properties of dynamically new comets with a perihelion less than 0.5 au. Such a detailed investigation of the activity evolution was carried out for the comet C/2012 S1 (ISON), the perihelion of which was 0.012 au; it was active at distances greater than 9 au from the Sun [40, 45, 55, 63, 106].

However, the number of comets observed in a large range of phase angles is rather small, and we therefore must observe a large number of different comets and draw conclusions from comprehensive investigations of various objects.

Since the spectra of most distant comets that we investigated did not exhibit gas emissions [42, 43, 60], we used broadband photometry to study the dust component.

The use of filters in photometry allows us to determine the color of the dust and dust productivity, study the active structures, measure the comet rotation period, and estimate the diameter of the comet nucleus. Using certain models gave us the possibility to estimate the dust particle size range, their velocities, and to localize the active regions. Below we describe the results of this analysis in detail.

### A.  Comet Dust Productivity Estimate

Determining the number of dust particles on the line of sight from the visible region continuum measurements is a rather complicated task, since the continuous light of the comet depends both on phase angle and on the distribution of particle by size, as well as their albedo and other factors which are little known to us in most cases, especially for LPC and distant comets. It also does not seem possible to separate the contribution to the luminosity of the comet of such parameters as particle albedo and their concentration. Nonetheless we can obtain an estimate of the dust productivity in comets from the scattered light flux of the comet in the continuum. The term "fill factor" has been used in comet physics for a relatively long time [4]: where $s$ is the cross section area of one dust particle and $N_d$ is the number of particles in a diaphragm of radius $\rho$. Since $sN_d(\rho)$ is the area of all dust particles observed in the diaphragm and $\pi\rho^2$ is the area of the diaphragm, in this case expression (??) represents the degree of filling of the diaphragm by comet coma dust particles as projected onto the celestial sphere. We thus derive the formula which is often used in comet physics for determining the relative measure of dust productivity in different comets [4]: where $Af\rho$ is a product of the Bond albedo ($A$), fill factor ($f$) and linear aperture radius ($\rho$), and the parameters $r$ and $\Delta$ are the heliocentric and geocentric distances to the object.

Since $f \propto N_d(\rho)\rho^2$ and $N_d(\rho) \propto 1/\rho$, the $Af\rho$ parameter does not depend on the size of the aperture. Formula (??) is widely used despite the fact that the condition of isotropic light scattering on the dust particles in comet atmospheres is clearly not fulfilled. However, this approach allows us to compare the results obtained in different epochs and with different instruments. But, despite the convenience of using this parameter, the results turn out to be strongly independent of the used apertures and the degree of comet activity when the condition of isotropic matter ejection from the comet nucleus is significantly disrupted.

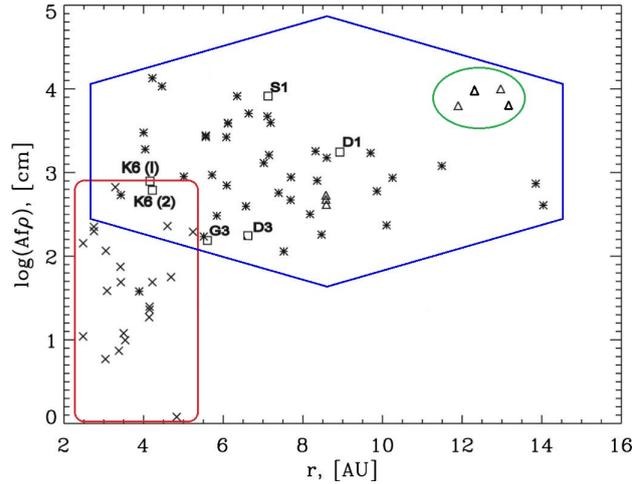

Figure 5: log ($Af\rho$) computed for the observed distant comets (squares) as a function of heliocentric distance $r$. For comparison, we show the parameter for LPCs from the literature (stars), for active centaurs (triangles), and for comets of the Jupiter family (crosses) which were active at large heliocentric distances [5, 40, 60, 65, 73, 78–85, 102, 109, 111, 113–115].

The $Af\rho$ parameter was computed for all comets studied with the 6-m telescope of SAO RAS (Fig. 5). This parameter turned out to be significantly larger for most of the distant comets than for ShPCs, including those of the Jupiter family observed at large heliocentric distances. Unfor- tunately, the sample of observed comets is still too small for a final conclusion, but a preliminary analysis shows a certain asymmetry in the activity, specifically: the post-perihelion activity of new comets is significantly higher than that at similar heliocentric distances, but before the perihelion. For example, an interesting result coherent with the dust component modeling was obtained for a distant comet C/2014 A4 (SONEAR) (hereafter 2014 A4). A conclusion about the fragmentation of particles in the coma, reached based on modeling, is confirmed by our photometric data (Fig. 6). A decrease in $Af\rho$ with distance from the nucleus can be most naturally explained precisely by particle fragmentation, which causes a decrease in the particle scattering cross section. Additionally, we obtained a significant difference in the $Af\rho$ profiles for the $r - sdss$ and $g - sdss$ bands. It is due to a change in the properties of light scattering by particles and confirms the dominance of sub-micron and micron sized particles in the coma of the distant comet.

A mass estimate of the dust ejected from the surface of the nucleus is often given in comet physics [43, 102]. All the techniques used for this purpose require knowledge of the particle albedo, their initial ejection velocity and size. Unfortunately, these parameters are not well known, especially for new comets. Space missions to short period comets, including the recent Rosetta mission to the 67P comet, have partially rectified this lack of information. However, most of the questions remain unresolved, and we can only introduce a given parameter as an estimate. As we showed in our paper Ivanova et al. [45], an accurate estimation of dust productivity from the $Af\rho$ parameter is an ambiguous task. The result of such a procedure depends strongly on the dynamic (e.g., effective density and cross section) and also optical (e.g., the scattering coefficient and phase function) characteristics of the dust particles. Slight deviations of the mentioned parameters can lead to drastic variations in the dust productivity estimate.

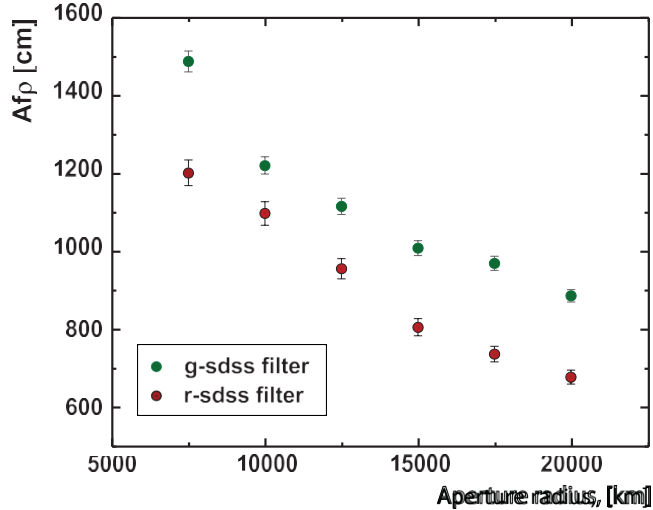

Figure 6: The *Afρ* parameter, measured in comet C/2014 A4 (SONEAR) in the *g sdss* and *r sdss* bands, as a function of aperture radius [42].

We have demonstrated that the microscopic properties of dust (effective size and porosity) link the dynamical and optical properties of comets with each other. One must also use the Mie scattering theory when computing the optical properties of porous dust particles (Fig. 7). One should also not forget about another uncertainty in the results of computations of the dust production rate for ejected dust—its chemical composition. The transition from the parameter *Afρ* to the mass value of dust productivity should be made with care, with account of the laboratory and space research for every individual comet.

### B. Dust Color Characteristics

Despite the uncertainty of the optical properties of dust particles [45], we can compute some dust characteristics. For example, the color analysis (estimate of the spectral reflective ability) often used in photometric observations shows (especially in aperture photometry) that the light scattered by dust particles in the coma is in most cases redder than the incident solar radiation. The spectral reflectivity of the comet $S(\lambda)$ at a wavelength $\lambda$ is determined as [53]: where $\varphi(\lambda)$ is the filter transmission curve, $Fs(\lambda)$ is the spectral distribution of solar radiation. Unfortunately, the efficiency of dust scattering $Q_{sca}(\lambda)$ cannot be measured directly from observations, however, it is assumed that for an optically thin atmosphere $S(\lambda) \propto Q_{sca}(\lambda)$.

For a quantitative characterization of dust scattering at different wavelengths, a spectral gra- dient of reflectivity is used [53]: where $dS/d\lambda$ is the reflectivity variation rate in the wavelength interval from $\lambda_1$ to $\lambda_2$, and $S'(\lambda_1, \lambda_2)$ is the average reflectivity in this wavelength interval.

Since the comet continuum is often observed with various filters, the spectral gradient of the re- flectivity is easier expressed in percentage at 1000 Å: where $\Delta\lambda$ is the effective wavelength difference between the red and blue filters expressed in Angstroms, and $F_1$ and $F_2$ are the comet radiation fluxes (in the case of aperture photometry) or intensities (in the case of surface photometry) in the red ($\lambda_2$) and blue ($\lambda_1$) spectral regions, correspondingly.

The characteristic property of reflectivity is that it depends directly on the micro-physical dust particle properties and does not depend on their amount in the field of view. For example, the mean values of normalized reflectivity gradients are within 3 to 25% [42, 46, 64] for the $B - V$ and $V - R$ spectral regions. If our aperture data are supplied by other color investigations from the literature [52, 79, 85], then the analysis results indicate that dynamically new comets ($1/a < 10^{-4}$ au$^{-1}$ [95]) that enter the inner region of the Solar System for the first time have slightly redder comas in the $B - V$ spectral region than post-perihelion comets. However, this conclusion must be confirmed with a large sample of new comets, especially considering the fact that aperture photometry averages the color, smoothing out the real picture of its distribution across the comet's coma (Fig. 8), as we can see in, e.g., the works of Ivanova et al. [42], Rosenbush et al. [101].

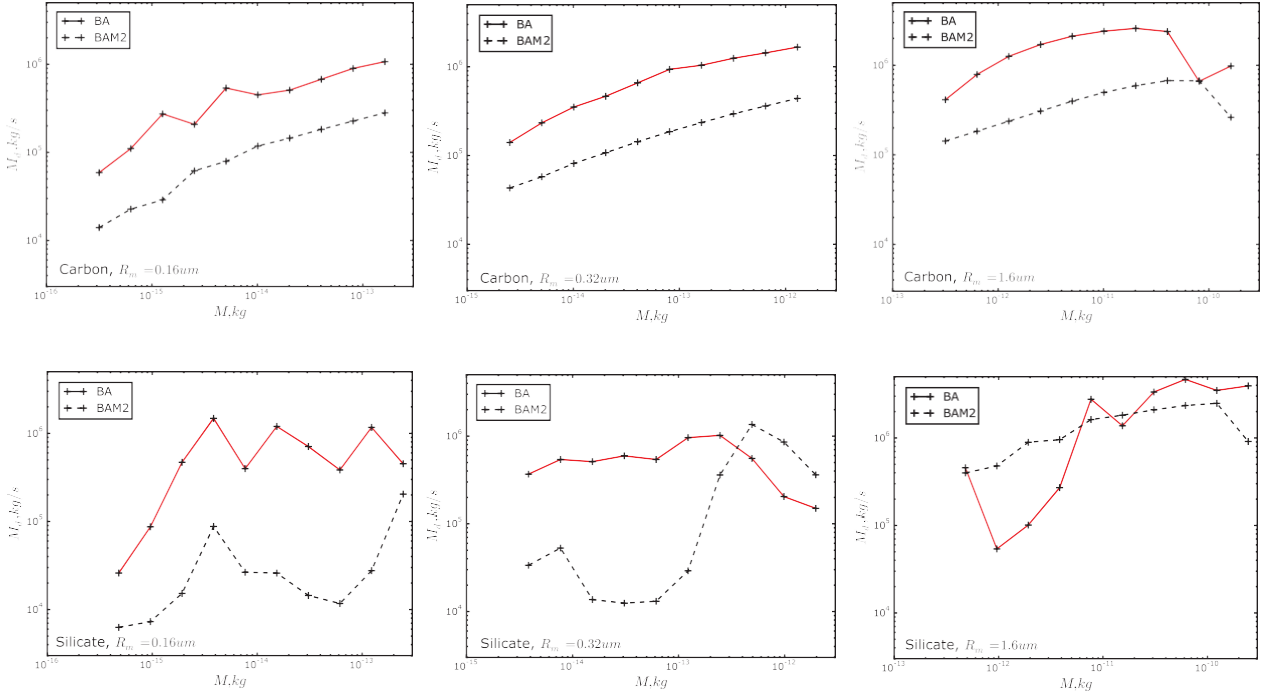

Figure 7: Computation results for dust productivity as a function of dust aggregate mass: top row—carbon aggregates, bottom row—silicate aggregates. The results are shown for three monomer sizes and two types of aggregates: BA and BAM2 [45].

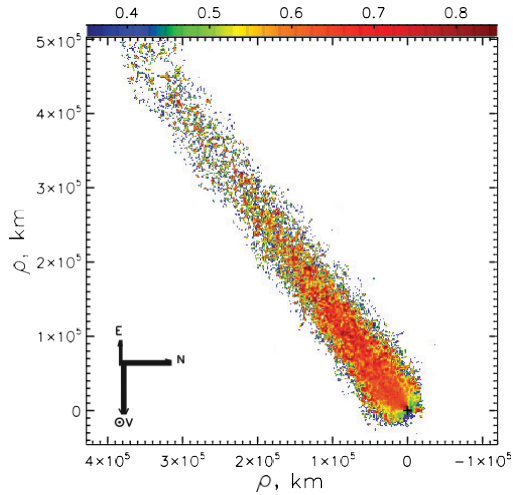

Figure 8: The $(g\ r)$-color map of the comet C/2014 A4 (SONEAR), constructed from the SAO RAS 6-m telescope images obtained on November 5, 2015 [42].

The question of dynamic color variations in the comas of distant comets, which we observed for several short period comets, also remains open [47, 74]. The main characteristic of photometric comet observations consists of the fact that for most comets, they are carried out for a limited number of times, and sometimes only once. The measurements are usually taken at large time intervals. The observed data are analyzed in the assumption of a certain degree of stability of the light scattering properties of comet dust. At the same time, it has been established that comets in general are unstable objects with various types of flare activity, accompanied by ejections of large amounts of matter. The question of how insensitive to micro-physical properties of comet dust are the characteristics of lights scattering, including the color index, remains open. It is important to understand whether the color index of a comet can undergo significant temporal variations which correlate with its flare activity, at least. For example, in the course of our studies Ivanova et al.

[47] we discovered that the dust color index in the comet C/2013 UQ4 (Catalina) was subject to significant short term variations. We suggested the most probable explanation of this phenomenon: these are temporal oscillations of the chemical composition of the dust particles in the inner coma. A search for fast variations of the color index is a new and relevant task in comet physics. Such a search was undertaken as far back as the 1980's within the framework of an international 1P/Halley comet research program. However, the research found neither significant temporal color variations nor even variations of linear polarization degree—a characteristic of scattered light that is most sensitive to micro-physical properties of dust particles [18]. This result was extrapolated to other comets and, thus, the task of searching for rapid variations in comets had lost its relevance for a long period of time. Nevertheless, various authors pointed out the presence of such variations due to the flare activity in comets (see, e.g., [28, 71, 117, 124]). There are some indications of fast color variations in the dust of comet C/1995 O1 (Hale–Bopp) [117] mentioned in the literature, as well as slow variations in the C/2013 A1 (Siding Spring) comet [71]. However, in the latter case, one cannot state with certainty that the variations were only of a long-term nature. The comet was observed with the Hubble Space Telescope and, as a result, the observation time was very limited; specifically, three observations were carried out over 4.5 months. We obtained new data on the short term color variations of the comet 41P/Tuttle–Giacobini–Kresak [74]. Its dust component modeling has shown that the chemical composition of dust in the inner coma changed overnight. We were also able to make conclusions regarding the composition of the inner coma of the comet, in particular, about the presence of silicates with a high concentration of magnesium and Mg–Fe organic compounds or silicates.

## C. $CO^+$ Ion Coma in Distant Comets

Our spectral data indicated a presence in two distant comets 29P and 2002 VQ94 of ions $CO^+$ and $N^+_2$ [42, 57–59]. This prompted us to investigate in more detail the ion coma of these comets and discuss the neutral gas ionization mechanism at large heliocentric distances. For this purpose, we selected the comet 29P. It is one of the most well-known among the few observed far comets at large distances (greater than 4 au) from the Sun, whose comas exhibit a presence of $CO^+$ ions [7, 9, 13, 49, 50, 58, 66]. Regular episodic flares are observed in 29P [39, 49, 97, 116]. The $CO^+$ ions have been detected both during flares and during the inactive state of the comet.

Various physical mechanisms were considered over the past several decades as sources of activity in distant comets [23, 25, 36, 37, 62, 120] and sources of ionization, including photoionization by solar radiation, collision ionization, recharging with solar wind protons, and also electric charge in the inner coma [14, 21, 34, 87]. Presently, there is no generally accepted view on which of the listed mechanisms is the most common. However, observations have shown that a significant part of comet ions form in the innermost coma (in the *collision zone*). Due to the low activity of such comets beyond the distances greater than 4 au, the solar wind may influence the surfaces of their nuclei directly, causing ion scattering of the surface material [6]. Since the ejection of comet material from the surface by scattering is practically proportional to the concentration of the components, this fact can be used to determine the composition of the nucleus. As the comet approaches the Sun, its coma becomes sufficiently thick for the solar wind to be absorbed above the surface, and the scattering ions disappear [91, 121].

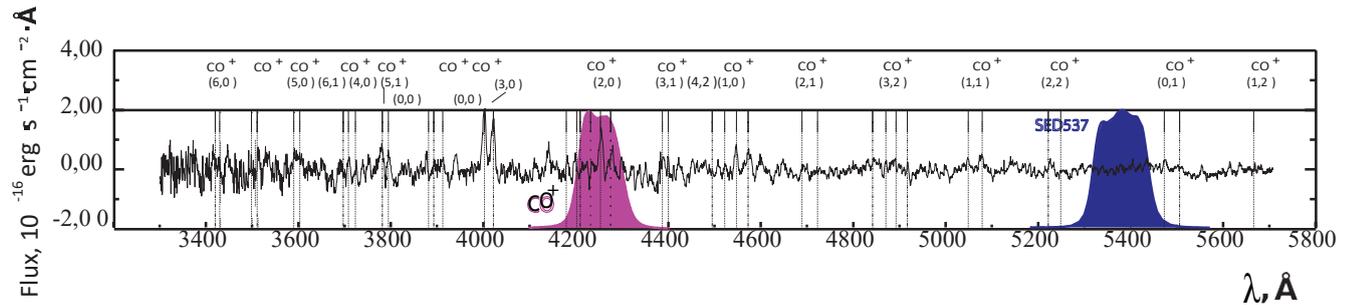

Figure 9: Normalized transmission curves of filters CO$^+$ ($\lambda$4277/39 Å) and *SED*537 ($\lambda$5309/169 Å), super- imposed on the observed spectrum of comet 29P [50].

The effects of ionization from the surface by scattering are known from the results of direct observations of comet 67P/Churyumov–Gerasimenko by the Rosetta mission [121] and from obser- vations in the vicinity of objects without any atmosphere. For example, Moon ions were scattered directly from the surface, and they are regularly observed in the solar wind as absorbing ions [29– 31]. In order to study the ionization mechanism at large heliocentric distances, we plotted and analyzed the CO$^+$ ion distribution maps in the coma of the comet 29P based on BTA observations carried out from 2007 to 2011 (see Ivanova et al. [42] and references therein). As we show in Fig. 9, the CO$^+$ ($\lambda$4277/39 Å) filter that we use lets through the emission of CO$^+$ ions and the continuum, whereas the *SED*537 ($\lambda$5309/169 Å) filter – only the dust continuum, without CO$^+$ ions.

To obtain a clean image of the ion coma of the comet we subtracted the summary continuum image, which was obtained in filter *SED*537, from the summary CO$^+$ image, taking into account the transmission curve of each filter and the difference in the continuum levels. We used the extinction values obtained during spectral observations of comet 29P [49, 50, 58]. The upper panel of Fig. 10 shows the CO$^+$ distribution maps after continuum subtraction, the lower panel shows the dust coma images of comet 29P. The CO$^+$ ion coma is observed on all dates. In most cases the coma is compact, asymmetric relative to the optocenter, and extended in the direction of the Sun. Unlike the ion coma, the dust coma was more stretched out and sufficiently symmetric. The comet coma transformed near the end of 2009: the activity of 29P diminished, and the dust coma became more compact. The CO$^+$ coma did not change significantly over this whole period, but was only slightly extended in the solar direction.

Based on an analysis of our observed data, we proposed a physical mechanism [39] explaining the CO ionization in the coma of comet 29P. Since we assume that the effects of surface ionization by scattering with charged particles of the solar wind play an important role (and even dominate) at a heliocentric distance of about 6 au, we analyzed the solar wind parameters in the vicinity of the comet and compared the solar wind particle flux with the observed CO$^+$ coma density, taking into account the fact that the ejection of matter from the surface during scattering is practically proportional to the flux. Direct measurements of the solar wind parameters in the vicinity of the comet nucleus are unavailable, and we used a numerical three-dimensional magnetohydrodynamic (MHD) model for the solar wind parameters in the heliosphere provided by NASA. This model is based on a numerical solution of the equations for plasma density, impulse, and energy density, as well as magnetic field in the MHD approximation [93, 94]. It was important for our purposes that the model covered ± 60deg in latitude, since the comet was positioned beyond the ecliptic plane. A comparison of the model with direct measurements at a distance of 1 au by space instruments STEREO-A, -B and Wind has demonstrated a relatively good agreement with the averaged parameters of the solar wind for the increased-density region (detected by all three space missions), through which the Earth passed on October 23–25, 2007, which thus verifies the local model results. This region of enhanced density was then spreading with the average solar wind velocity (of the order of 380–430 km s$^{-1}$).

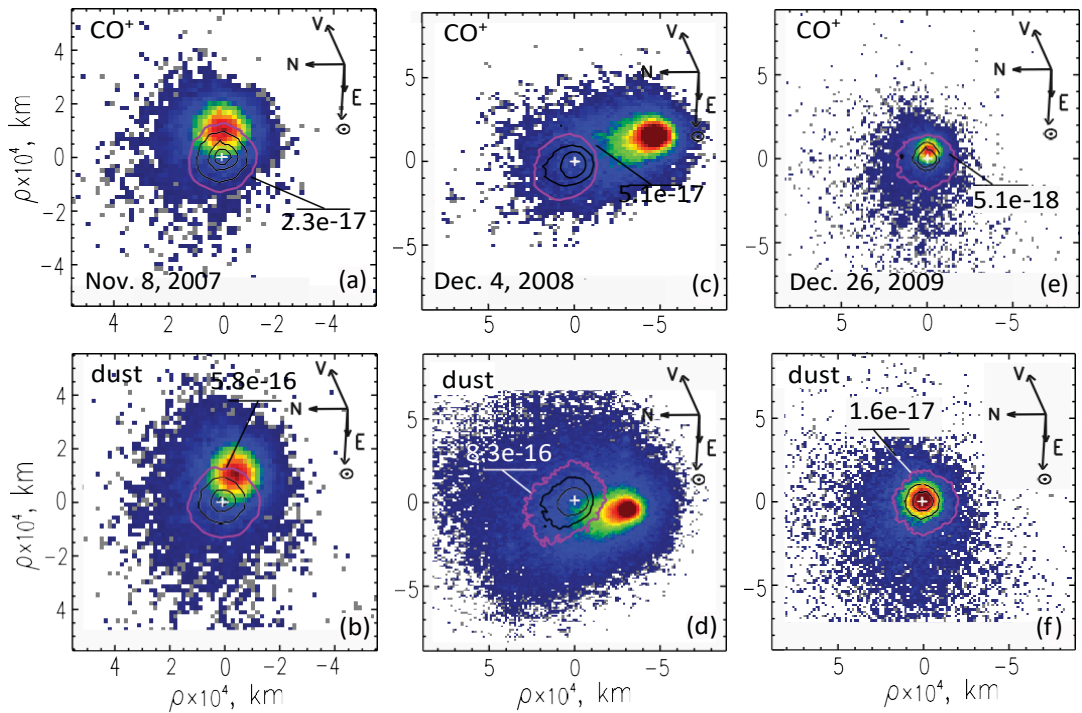

Figure 10: Isolated CO$^+$ ((a), (b), (c)) and dust ((d), (e), (f)) comas of comet 29P, obtained from the 2007 observations with the 6-m telescope of SAO RAS. All the images are scaled in erg cm$^{-2}$ s$^{-1}$ Å$^{-1}$ [50].

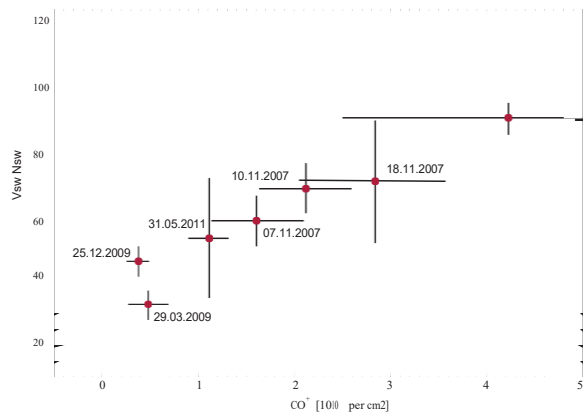

Figure 11: Influence of the solar wind proton flux on the CO$^+$ ion flux in the vicinity of comet 29P [39].

An almost linear dependence (see Fig. **??**) indicates that scattering ionization, which is assumed to be proportional to the particle flux, is the most effective ionization mechanism for comets at large heliocentric distances (about 6 au for 29P).

### D. Morphology of Active Structures in Selected Distant Comets

Most of the small bodies in the Solar System exhibiting comet activity are seen to have various structures in their comas: jets, fans, envelopes, etc. These essentially spatial variations of coma density result from processes related to the nucleus or the interaction of a comet with solar wind. Active structures can be used to study the properties of the nucleus, invisible with optical tech- niques. Additionally, the morphology carries information about the variations which took place during the formation of the active structures. For example, periodic variations and features of the coma can be used to estimate the rotation period of the comet nucleus.

The importance of morphological studies in comet physics consists of the detection of spatial structures which are visible in almost all comets independently of their dynamical class, and almost at all wavelengths. The structures can be formed by dust, ions, neutral gas. Depending on the used photometric filters we can isolate structures in the comet coma that are formed by one of the mentioned components of the comet atmosphere. The structures are often seen both in the near-nuclear regions and up to distances greater than tens of thousands of kilometers from the nucleus.

The structures can be both long-term and stable formations, observed for up to a month and longer, and short-term, with lifespans of several minutes. Individual active features in the comet comas can be both periodic and of spontaneous nature. The same is true for their shapes, which are diverse and vary from symmetric to completely chaotic. In our studies of distant comets we analyzed the morphology of active structures using photometric data. Since most far comets do not demonstrate a presence of strong gas emissions or ions in their spectra (see the section on spectral studies), broadband filters are used to study the morphology. In most distant comets that we investigated, the coma and distant structures were formed by dust particles ejected from the nucleus. We used narrow band filters only in observations of two comets—200 VQ94 and 29P, in order to isolate and study the dust and $CO^+$-comas separately [38, 58]. Most of the distant comets exhibit activity with the presence of active structures and extended tails. On the outside, they do not resemble the comet tails (Fig. 12) that form at close heliocentric distances. They usually do not have an internal structure, display an almost constant width along the tail, and are often strongly bent.

In order to isolate low-contrast structures in the comet coma, we used several image analysis methods: dividing by the azimuth average, azimuth renormalization [104], and also the rotational gradient method [67]. The first two methods allow one to compensate the bright comet coma background and isolate the low-contrast structures. Dividing by the azimuth average works very well for isolating bright broad jets. All the images were checked for artifacts. Image reduction by digital filters was mainly carried out using two different programs: one from the Planetary Space Science[4] website, and our own IDL codes for all of these filters. As we mentioned above, the intensity maps were constructed by combining all photometric images obtained with the same filter.

Since different filtration methods affect the image differently and can also lead to an appearance of false structures after the image filtering, we used each of the digital filters on an individual frame, and then on the summary image to estimate whether the detected structures are real or not. This method allowed us to avoid the appearance of false structures during image reduction using digital filters (Fig. 13). Additionally, we also investigated the structure variation due to a shift of the comet optocenter.

Such a technique of isolating low-contrast structures was successfully used for various distant

---

[4] http://www.psi.edu/research/cometimen

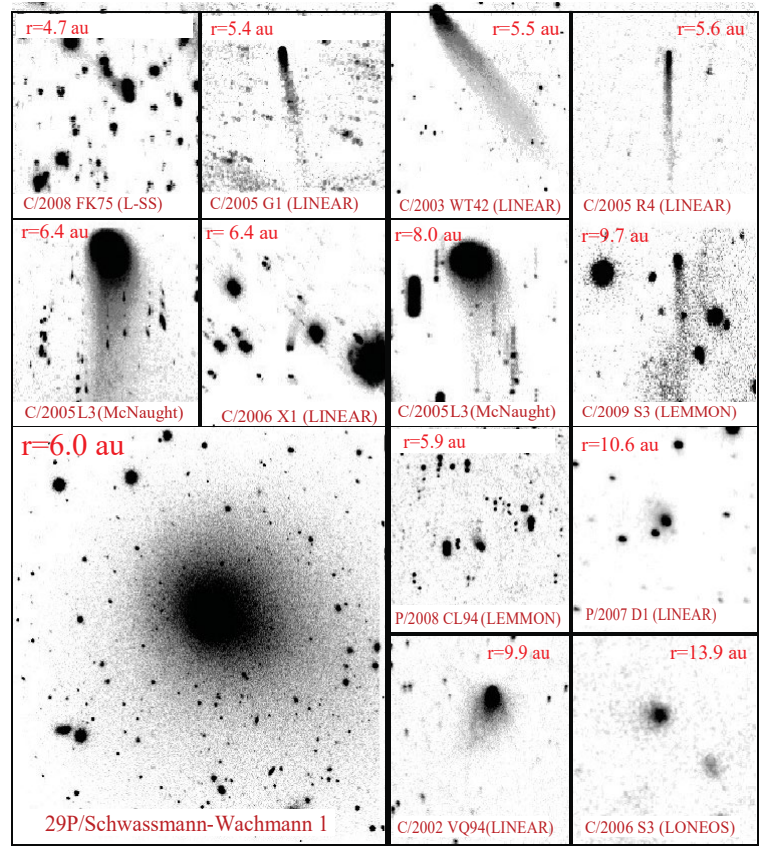

Figure 12: Direct images of selected distant comets with various morphological manifestations, obtained with the 6-m telescope of SAO RAS during the 2006–2011 period.

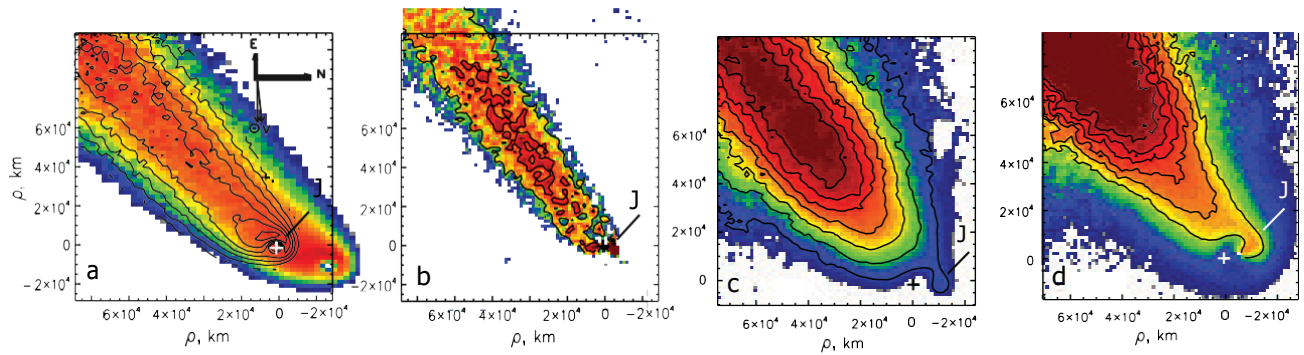

Figure 13: *r sdss* band images of comet C/2014 A4 (SONEAR) obtained with the 6-m BTA telescope of SAO RAS [42], reduced by means of digital filters: (a)—relative intensity image, reduced using the division by the $1/\rho$ profile [104]; (b)—image reduced by the rotational gradient method [67]; (c) and (d)—relative intensity images reduced by means of azimuth average division and renormalization [104]. $\rho$ is the distance from the comet optocenter, *J* is the fan-like structure in the circumnuclear region. The arrows indicate the direction towards the Sun, the north (N), the east (E), and the comet velocity vector projected onto the sky plane ($V$).

comets [38, 42, 47, 58, 60, 101].

### E.   Rotation Period Estimate for Distant Comets with Jet Activity

Estimating the comet rotation period is an important task for understanding and describing the physical processes that are responsible for comet nucleus activity. To date, this also remains relevant for distant comets. Different methods can be found in the literature for estimating the rotation period [105]. Most of the described estimates of the rotation period are based on the analysis of comet light curves obtained with broadband filters. Also, along with the rotation period determination from comet light curves, a method based on studies of the dynamics of active structures in comets is widely used. Different versions of this method were used earlier for investigations of comets 29P [112], C/1995 O1 (Hale–Bopp) [72, 105], P/1992t (Swift–Tuttle) [123] and others. We used this method for investigating the rotation period of the nucleus of comet 29P, which is known for its flare activity with the formation of multiple structures in the coma. To find the period, we proposed a modified method which, unlike the other techniques, is almost independent of external parameters (gas velocity, comet magnitude error, etc.), since it is based on the estimate of the shift of low-contrast structures in the images. The only allowance in our method is the assumption that we register the angular shift of the same structures during the period of observations.

To verify our results we compared them to similar results obtained from observations carried out at different time periods with the 6-m SAO RAS telescope and the 2-m telescope of the Mount Terskol Observatory of the International Center for Astronomical, Medical and Ecological Research. In order to isolate the low-contrast structures (jets) in the images of the dust coma of comet 29P, we used the method described in Ivanova et al. [38] of using digital filters to reduce images. As is evident from Fig. 14, the use of digital filters on the 29P comet images obtained in December 2008 reveals two jets which were visible during the entire observing period.

A cross-correlation method was used to analyze the angular shifts of the structures isolated in the comet coma. The images were converted from Cartesian coordinates to a polar coordinate system with the center corresponding to the photometric center of the comet. The radial distance from the nucleus was selected as the polar distance, and the azimuth angle, counted anticlockwise from the north direction, as the polar angle. Additionally, prior to the coordinate system conversion, all the images were similarly aligned in the direction to the north and east. To avoid possible erroneous results in determining the comet rotation period, the low frequency trend was removed from the comet images that were not digitally filtered. As a result, we isolated two jets in the images obtained in 2008, and three dust structures in the images from 2009. To verify our rotation period estimate, we had to ensure that the isolated structures in the comet coma were the same during the entire observing period in each of the sets. In our case, we had two sets: in December 2008 and in February 2009. In order to verify our assumption, we used the synchronous detection method.

The estimate of the period for the first series of observations (December 3–5, 2008) amounted to $12.1 \pm 1.2$ days, and for the second series (February 14–19, 2009 г.) to $11.7 \pm 1.5$ days. These estimates do not contradict each other, and the slight scatter in the results may be explained by
the measurement errors. The results are close in value for different observing periods and show that the used method is rather accurate and can be used in observations of other distant comets with active structures in the dust coma.

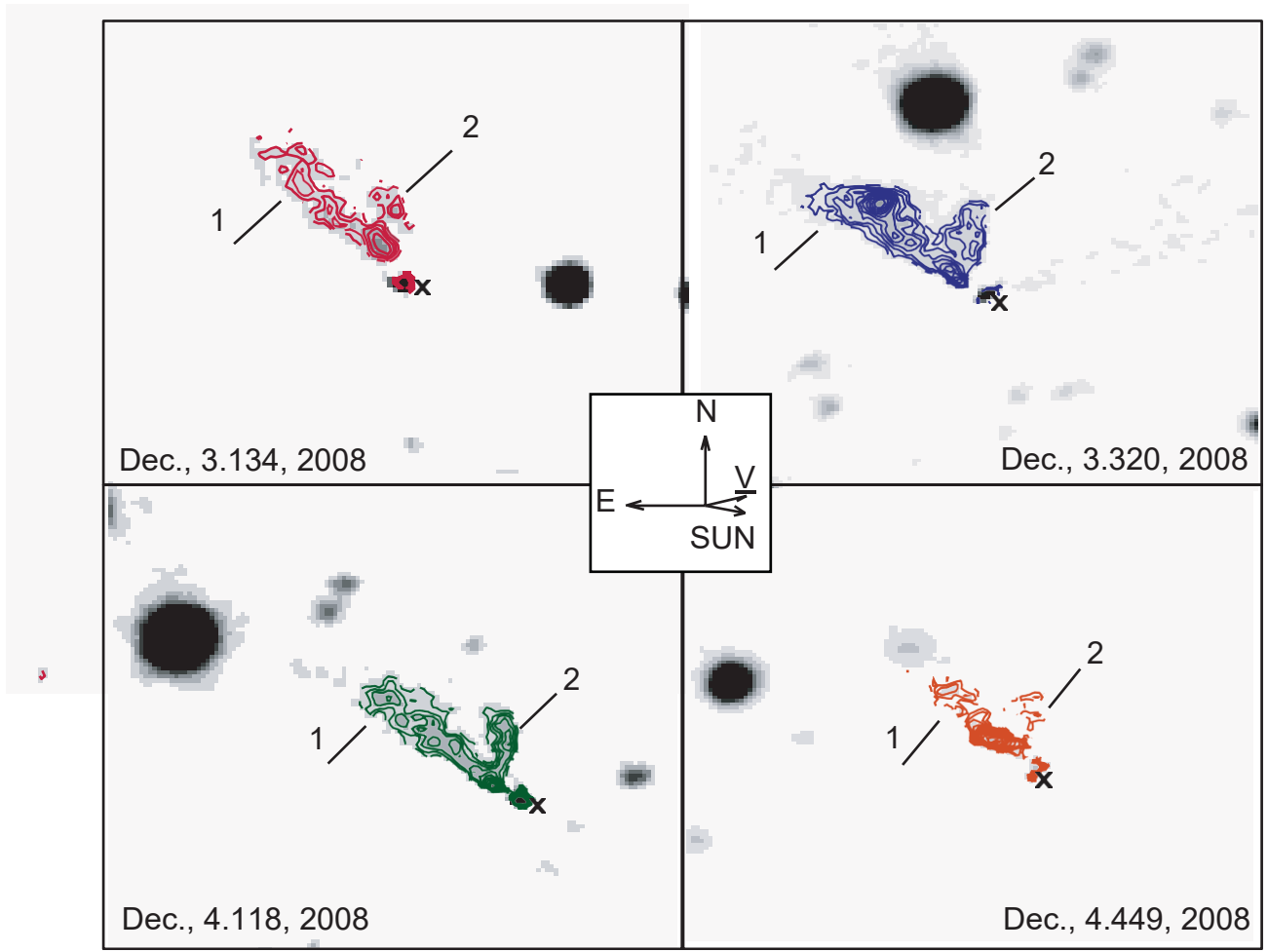

Figure 14: Isolated structures *1* and *2* in the dust coma of comet 29P after the reduction of the summary images of the comet, obtained in December 2008 with the 6-m telescope of SAO RAS, by the rotational gradient method [67]. The cross marks the position of the comet's photometric center [38].

## IV.  POLARIMETRIC OBSERVATIONS OF SELECTED DISTANT COMETS

Most of our knowledge of the physical properties of comet dust was obtained from comet observa- tions in the moments when they approach the Sun (less than 2 au). Also, prior to our investigation, the nature of the dust particles in the comet coma was considered to be independent of heliocentric distance [18]. Nevertheless, new research shows the differences between the activity and physical properties of comets close to the Sun and those that are distant [17, 43, 51, 58, 59, 82, 85]. It is therefore important to accumulate observational material and to understand from analysis why in some comets the dust is represented mostly by compact particles, while in others porous aggre- gate particles are dominant. To this end, one must analyze the connection between the dynamic characteristics of comets and their physical properties. Also, one of the reasons for the difference in ShPCs and LPCs may be in the comet evolution. ShPCs, unlike the new and long-period ones, have undergone multiple passages near the Sun, which can reflect significantly on their physical properties. In 2011, the 6-m telescope of SAO RAS started a new program aimed at investigating the properties of dust in distant comets, which are active in the region of the Solar System where the sublimation of water ice is insignificant. This task was relevant, and it seamlessly supplemented the previous results obtained by spectrophotometric methods of studying distant comets.

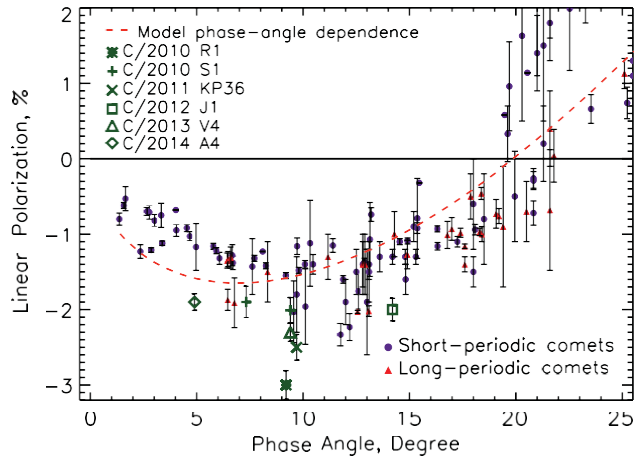

Figure 15: Summary phase curve for short-period and LPCs at small heliocentric distances and for distant comets (the symbols are shown in green) at large heliocentric distances [17, 48].

We must note that at that time, practically no data were available on the polarization of comets that were active beyond the orbit of Jupiter. Polarimetric studies that were carried out for six years allowed us to obtain new data on the size of the dust particles in selected distant comets, and make conclusions regarding their orientation and composition. Since some of the distant comets are characterized by flare activity, like, e.g., comet 29P, we investigated the variations of the scattering properties of dust particles during the times of flares. An investigation of spectral dependence of the polarization of light from distant comets allowed us to make conclusions about the dust properties, including their fragmentation. The first such analysis of linear polarization distribution in the comet coma for distant comets with a perihelion greater than 4 au was published in the works of Ivanova et al. [46, 48]. All the results of our observations conducted at phase angles $\alpha < 15$ deg show that the negative polarization branch is significantly more pronounced (Fig. 15, the data are taken from the NASA PDS website[5]) than for comets at small heliocentric distances. For example, the polarization of comets C/2010 S1 (LINEAR) (hereafter 2010 S1) and C/2010 R1 (LINEAR) (hereafter 2010 R1) is significantly higher (polarization degrees in the coma range from −1.9% to −8%) than the typical polarization (−1.5%) observed for most comets at close heliocentric distances. The results of numerical modeling [48] using the $T$-matrix superposition method [76] has shown that a model of comet dust in the form of aggregates with a common radius of about 1.3 microns, porousness $a = 0.54$, and refraction index $1.65 + i0.05$ works well for comets 2010 S1 and 2010 R1.

Other results were obtained for an intermediate-type comet (where water and $CO_2$-ice compete for dust ejection) C/2012 J1 (Catalina) (hereafter 2012 J1). Taking into account the distribution maps and profile analysis of the linear polarization degree in the comet coma, it is evident that

5 https://pdssbn.astro.umd.edu/holdings/pds4-compil-comet:polarimetry-v1.0/SUPPORT/dataset.html

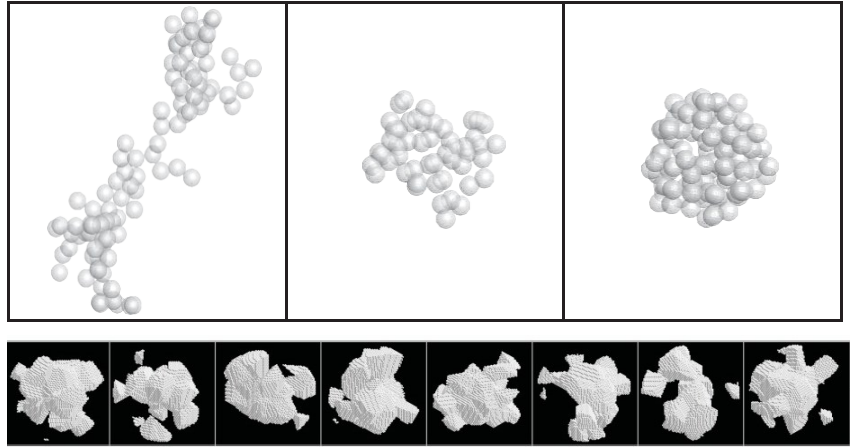

Figure 16: And example of dust particles used in distant comet modeling: upper panel—particles in the form of aggregates [48], lower panel—particles in the form of agglomerated fragments [47].

the coma of comet 2012 J1 does not have any morphological peculiarities [48]. In other words, the degree of linear polarization is almost constant, there are no areas of negative polarization, and therefore, we can assume that the dust coma of the comet consists of dust particles that have the same optical characteristics.

The average linear polarization degree in comet 2012 J1 is determined at the level −2%. Observational data modeling, based on a simple approach when one type of particles was taken with the same refractive index and the same particle size distribution in the entire coma, served as a demonstration of the fact that such a polarization level can be produced by all materials with the imaginary component of the refractive index less than 0.07. ALso, the spectral gradient from observations introduced additional conditions on the particle composition, specifically: one cannot describe the observed degree of polarization by the influence of water ice, the presence of which is expected in comet 2012 J1. The authors [46] determined that this degree of linear polarization and this spectral gradient are simultaneously possible for matter with a refractive index of $1.6 + 0.03i$, which corresponds to silicates enriched with magnesium and a small abundance of amorphous carbon.

Polarimetry analysis for six distant comets allowed us to combine the data and obtain theoretical modeling results for light scattering properties for particles of different morphology, and also discuss based on the results of this comparison the possible particle composition in the atmospheres of distant comets. In particular, we considered aggregate particles (both very small and very porous) consisting of a large number of sub-micron spherical monomers (Fig. 16).

As a result of numerical modeling based on the $T$-matrix method [76], we obtained a reasonable agreement with all the observed polarimetric data for the following two particle models: a mix- ture of flattened water–ice spheroids with porous aggregates, consisting of silicate monomers, or compact aggregates, consisting of silicate monomers; and, a mixture of the same flattened water– ice spheroids with porous silicates and organic aggregates. We should note the one of the main differences between our model and those used for comets located near the Sun is the inclusion in the model of a large number of particles consisting of water ice. A comparison of theoretical and observed data has shown that the presence of ice particles (in this case, in the shape of flattened spheroids) leads to a deepening of the negative polarization branch and thus improves the agree- ment with the observed data for distant comets. The latest detailed investigation of a distant comet 2014 A4 [42] confirms our previous results which demonstrate that for distant comets, the negative polarization branch at small phase angles is deeper compared to the ones observed for comets near the Sun (Fig. 17).

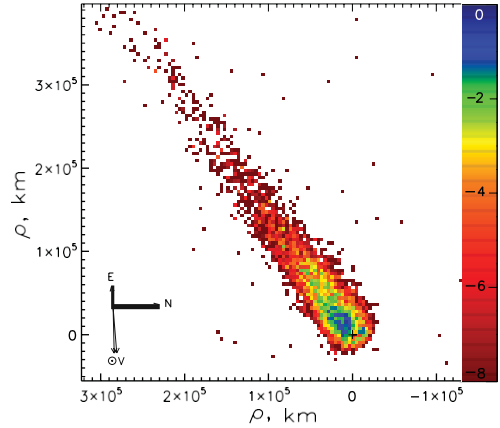

Figure 17: Distribution of the degree of linear polarization in the coma of the C/2014 A4 (SONEAR) comet. Polarimetric images were obtained in the $R$-band with the 6-m telescope of SAO RAS [42].

The only case of a large negative polarization ($-6\%$ in the halo of comets 81P/Wild 2 and 22P/Kopff) was reported in Hadamcik and Levasseur-Regourd [27]. However, these comets were observed at the phase angles of 97 and 18 deg. At such phases angles, the polarization can reach significant negative values even for a sufficiently orderly dust composition. For comet C/2012 S1 (ISON), the average (negative) polarization in the coma was found to be $-1.6\%$ at heliocentric distances of 3.81 au [32]. Such a negative polarization is unusually low in comparison with those that are observed for other distant comets and is possibly related to the uniqueness of comet C/2012 S1 (ISON), which turned out to be a sungrazer.

In our observations of comet 2014 A4 the linear polarization distribution map shows spatial variations of polarization in the coma ranging from about $-3\%$ near the nucleus to almost $-8\%$ in the tail (Fig. 13). An analysis of polarization, color, and their variations with distance from the nucleus shows that the dust particles in distant comets are very small (less than 1 micron), which is unsurprising, since there is not enough gas to lift larger particles. The dust contains ice particles, as well as particles consisting of (or covered with) tholin or other similar organics. Ice particles get fragmented as they recede from the nucleus.

Photometric studies of comets confirm the fact that the coma is formed by sub-micron and micron particles, which become fragmented with time. Thus, our results show that the dust in distant comets differs from the dust in regular comets, which are observed at heliocentric distances less than 2 au, both in size and in composition. We must note the polarimetric data for comets close to the Sun are, usually, well described by a mixture of rather large particles made out of dark material and silicates [56].

## V.   ACTIVITY MECHANISMS AT LARGE HELIOCENTRIC DISTANCES

Although the physical activity of comets beyond the water ice sublimation zone (greater than 3 au) was registered a long time ago [100], to date, the mechanisms of its formation and evolution are still not thoroughly understood or well-founded. Various comet activity mechanisms at large heliocentric distances are proposed in the literature:

• sublimation of supervolatile $CO/CO_2$ ices [33, 35];

• phase transition of primitive amorphus ices, of which the comet nuclei generally consist, into a

crystal form [16, 26, 98];

- HCN polymerization; [99];

- meteoroid bombardment of the comet nucleus surface, which leads to the activation of other mechanisms [44, 110];

- melting of subsurface ices and formation of supervolatile component solutions, which is accom- panied by significant energy release [86, 88];

However, most of the proposed mechanisms cannot explain the observed long-term activity of comets at large heliocentric distances [51] and require further investigation and accumulation of new material. An uncertainty also remains in the chemical composition of the atmospheres of distant comets, their gas and dust productivity, structure and constitution of the nuclei, which  is directly related to comet cosmogony. In our opinion, revealing the differences in the physical properties of comets belonging to different dynamical types will allow us to solve the problems of both the homogeneity of the nuclei composition, and the homogeneity of matter in the comet formation region in the protosolar nebula. No less important is the use of a large amount of accumulated data to compare the general observed physical properties and individual features of comets with their dynamic characteristics, which point to their potential places of origin.

## VI.   OPEN QUESTIONS IN THE PHYSICS OF DISTANT COMETS

The uniqueness of the comprehensive observations carried out with the 6-m telescope of SAO RAS equipped with the SCORPIO and SCORPIO-2 instruments consists of obtaining a large volume of data by various methods, which allows not only a more detailed study of the dust and gas properties in selected objects, but also the introduction of restrictions on model parameters that are used to interpret the observed data. This approach allowed us not only to collect unique data for a new type of comets, but also to improve our understanding of the evolution of dust in the coma, supplementing the color and polarization analysis by a study of variations of other photometric characteristics of dust, specifically, by radial profiles and variations in dust productivity with distance from the nucleus. Spectral studies allow us to consider the presence of supervolatile components in distant comets, which could be responsible for the formation of their extended comas and tails.

Undoubtedly, the obtained models for dust particles in the atmospheres of a small number of distant comets are preliminary, although they allowed us to place restrictions on the composition and size of the particles. It is also important to remember that each comet is unique and, unfor- tunately, one can rarely trace the evolution of LPCs or obtain a long series of observations, which would include the period before and after the perihelion passage. Oftentimes we obtain episodic observations, which relate to no more than a few phase angles per comet. Therefore, for a better understanding of the nature and morphology of particles in the atmospheres of distant comets, future additional comprehensive observations of comets are required, both photopolarimetric and spectral, conducted at different phase angles and with the use of several filters.

The question of the presence of circular polarization in distant comets remains open. It was detected and studied with the 6-m telescope in the /2009 1 (Garradd) comet [38]. Its evolution with distance from the Sun was studied at close heliocentric distances from 1.6 to 2.2. au. New observations are needed of distant comets with significant activity at distances greater than 4 au,  allowing us to register a weak circular polarization signal.  Looking into this task can help  us understand whether circular polarization is more typical for new comets as opposed to old short-period ones, and how it depends on the evolutionary processes.

The presence of silicate peaks in distant comets is also not well studied. The spectra of some comets in the infrared spectral range (IR) show silicate emission bands above the continuous spectrum, hinting at a complex mineralogy of the comet silicates [108]. The strongest silicate emissions were observed in the C/1995 O1 (Hale–Bopp) comet, although they were very similar in shape to those that were observed in Halley's comet and others [107]. The intensity of the silicate peaks depends on the size and temperature of the particles. Not all comets exhibit strong ten-micron emission peaks. Practically all ShPCs (with the exception of the disrupted 73P comet) exhibit no strong emission peaks, and some of them lack emissions in the IR region altogether.

Another unresolved question is that of the birth place of comet 29P, which, based on optical and radio data, is enriched with and, correspondingly, should have formed in the cold outer regions of the protoplanetary disc [96]. Nonetheless, the latest data from the Spitzer telescope show a presence of crystal silicates in the comet coma [54, 112], which underwent severe thermal processing in the regions near the young Sun. These contradictory observational data can indicate a possible migration of the comet to a hotter area during its dynamic evolution. Our research of the dynamic evolution of the orbit of comet 29P [89] has shown that the low inclination of the comet's orbit is more indicative of its origin in the trans-Neptunian region. Since the comet has most likely migrated through a large interval of heliocentric distances, from a position deep inside the orbit of Mercury to the trans-Neptunian region, its surface can no longer be considered as pristine, as it was probably already altered by the interaction with intense solar radiation and interplanetary medium.

Laboratory studies show that the evolution of activity has typical features depending on which physical mechanism caused it [92]. Our theoretical investigations [51] have shown that the mecha- nisms presented in the literature (see section IV.A) soundly describe the temporary flare activity in comets at large heliocentric distances, but cannot explain the long-term activity that we observe in distant comets when a large volume of matter is ejected from the surface [51]. No less impor- tant is the question of interrelation of objects exhibiting comet activity, but differing in dynamic properties in a wide range of heliocentric distances. For example, to date, there are practically no polarimetric studies of centaurs and active asteroids.

The obtained results of distant comet studies show that despite the general similarity in the properties of comets from different dynamic groups, there are various deviations in their individual characteristics, which are mainly determined by the differences related to the location of their formation and consequent evolution, including migration in the Solar System under the influence of giant planets.

Comprehensive observations of distant comets with the 6-m SAO RAS telescope have signifi- cantly improved our knowledge in the field of distant comet physics, which is especially important in advance of the new ESA mission "Comet Interceptor" (the first mission to a LPC). However, despite the more than a ten year period of distant comet investigations with the 6-m telescope of SAO RAS, the accumulated data are still not sufficient for a full scale analysis and extrapolation to the cases of other distant comets. We need to continue specifically with our comprehensive studies using different methods, including photometry, polarimetry, and spectroscopy. This will help us to further organize the data and isolate the taxonomic SSSB classes which exhibit comet activity, and understand the influence of the place of origin in the Solar System on the composition and properties of individual comets and groups.

### Acknowledgments

Most of the works have been carried out within the framework of comprehensive observa- tional programs for research of distant comet activity together with co-authors in SAO RAS (Afanasiev V. L., Moiseev A. V.) and MAO NASU (Korsun P. P., Rosenbush V. K., Kiselyev N. N., Kulyk I. V.) The author would like to thank Alexey Moiseev and Igor Lukyanyk for critical com- ments and discussion, which have significantly improved the paper. The author is grateful to Afanasiev V. L., Moiseev A. V., Oparin D. V., Uklein R. I for conducting observations with the 6-m telescope of SAO RAS.

## FUNDING

Observations with the SAO RAS telescopes are supported by the Ministry of Science and Higher Education of the Russian Federation (including agreement No. 05.619.21.0016, project ID RFMEFI61919X0016). The research was additionally funded by a grant from the Slovak academy of sciences VEGA 2/0023/18.

---